\documentclass[12pt]{article}
\usepackage[cp1251]{inputenc}
\usepackage{amsmath}
\usepackage{eufrak}
\usepackage{euscript}

\makeatletter

\renewcommand{\section}{\@startsection{section}{1}%
{\parindent}{4.0ex plus 0.8ex minus .2 ex} {1.6 ex plus .2 ex}{\large\bf}}   % \large
\makeatother

\begin{document}

\title{On Field Theory and Some Finsler Spaces}

\author{G.I. Garas'ko \\[6pt]
%EndAName
 Russia Electrotechnical Institute, Moscow, Russia\\[6pt]
  gri9z@mail.ru}

  \maketitle

\begin{abstract}
The paper introduces the approach to construction of the Lagrangian of the
field (fields). This approach is based solely on the metric function of the
Finsler space: the Lagrangian is constructed as the unit divided by the
volume swept by the unit vector running through all the points of the
indicatrix in the tangent space under the assumption of the tangent space
being Euclidean. For the space, which is conformally connected to the
Minkowski space, under the assumption of the exponential time dependence and
spherically symmetrical coordinates dependence the cosmological equation is
written, which yields Hubble law for distances from the origin which are
much less than the size of the universe. The cosmological equation is
written for the field describing the universe with the geometry conformally
connected to the geometry of polynumbers H(4) with the Berwald-Moore metrics.
\end{abstract}

\section{Introduction}

Both in classical theory \cite{1} and in the theory of quantized fields \cite{2} the
most `convenient' method of field equations construction deals with such concepts as
Lagrangian, action and the principle of the least action (Hamilton's principle).
According to this approach, the relation is defined unambiguously \cite{3} between
continuous transformations (with respect to which the action is invariant) and
physical laws of conservation, that can be verified empirically.

If $x^{0},x^{1},x^{2},x^{3}$ are coordinates, $f(x)\equiv
f(x^{0},x^{1},x^{2},x^{3})$ is a scalar field in Minkowski space, and $\mathfrak{L\
}$, given by
\begin{equation}
\mathfrak{L}\equiv \mathfrak{L}\left( f(x);\frac{\partial f}{\partial x^{0}},%
\frac{\partial f}{\partial x^{1}},\frac{\partial f}{\partial x^{2}},\frac{%
\partial f}{\partial x^{3}}\right) \,,  \label{1}
\end{equation}
is the Lagrangian, then the integral of the Lagrangian over certain
4-dimensional volume $V$ in space-time,
\begin{equation}
I[f]=\int\limits_{V}^{\quad 4}\mathfrak{L}\,\,dx^{0}dx^{1}dx^{2}dx^{3}
\label{2}
\end{equation}
is said to be `action'. Under the assumption, that variations of the field
function $\delta f$ are equal to zero on the boundary of integration domain,
and taking into consideration the requirement of stationarity of action,
\begin{equation}
\delta I[f]=0\,,  \label{3}
\end{equation}
applying the well-known method, we get Euler-Lagrange equation, the field
equation:
\begin{equation}
\frac{\partial }{\partial x^{i}}\frac{\partial \mathfrak{L}}{\partial \left(
\frac{\partial f}{\partial x^{i}}\right) }-\frac{\partial \mathfrak{L}}{%
\partial f}=0\,.  \label{4}
\end{equation}

Usually Lagrangian is selected with the purpose to obtain finally the given field
equations, or constructed with the purpose to ensure the desired symmetry and to
meet certain auxiliary requirements: e.g. when selecting the Lagrangian we may try
to obtain the linear partial differential equations of second order. Construction of
the essentially new Lagrangians describing non-linear physical processes reperesents
is, in certain sense,  a kind of `art'.

The functional (\ref{2}) may be interpreted from a purely
geometrical standpoint:\ i.e. not as the integral (of the Lagrangian $%
\mathfrak{L}$ as the integrand) in the Minkowski space, but as the volume in
the space (more complex space)\ with the volume element given by:
\begin{equation}
dV=\mathfrak{L}\,\,dx^{0}dx^{1}dx^{2}dx^{3}\,.  \label{5}
\end{equation}

Consider the Finsler space $x^1,x^2,...,x^n$ \cite{4} with the metric function
\begin{equation}
L(dx;x)\equiv L(dx^1,dx^2,...,dx^n;x^1,x^2,...,x^n)\,.  \label{6}
\end{equation}
In this space, let the length element $ds$ be defined by
\begin{equation}
ds=L(dx^1,dx^2,...,dx^n;x^1,x^2,...,x^n)\,.  \label{7}
\end{equation}
The metric properties of Finsler space may be more evidently described in
terms of  {the }concept of indicatrix. In every point $%
M(x^1,x^2,...,x^n)$ of the main space the indicatrix is defined in the
corresponding tangent centroaffine space $\xi ^1,\xi ^2,...,\xi ^n$ as a
hyperspace made up from the `endpoints' of unit radius-vectors $\xi _{(1)}$.
Points of this hypersurface satisfy the equality:
\begin{equation}
L(\xi ^1,\xi ^2,...,\xi ^n;x^1,x^2,...,x^n)=1\,.  \label{8}
\end{equation}
If the system of indicatrices is defined in every point of the main space,
or (what is the same) the sets of unit vectors are defined, the Finsler
geometry is defined completely. To calculate the length of the vector $%
(dx^1,dx^2,...,dx^n)$,  {one has }to find a unit vector $\xi _{(1)}$ co-directional
with the vector $dx$, then the scalar coefficient $ds$ in the relation
\begin{equation}
dx^i=ds\cdot \xi _{(1)}^i  \label{9}
\end{equation}
will be the length of the vector $dx$. From the last relation it follows
that, the length element
\begin{equation}
ds=\frac{|dx|_{\hbox{\small{eu}}}}{|\xi _{(1)}|_{\hbox{\small{eu}}}}\,,
\label{10}
\end{equation}
where $|dx|_{\hbox{\small{eu}}}$, $|\xi _{(1)}|_{\hbox{\small{eu}}}$ are lengths of
vectors $(dx^1,dx^2,...,dx^n)$ and $(\xi ^1,\xi ^2,...,\xi ^n)$ respectively,
calculated as if the spaces $dx^1,dx^2,...,dx^n$ and $\xi ^1,\xi ^2,...,\xi ^n$ were
Euclidean, and coordinate systems employed were Cartesian.

If under these assumptions, this is possible to calculate the volume of the
indicatrix, i.e. the $n$-dimensional volume, swept by the unit vector $\xi _{(1)}$
in the tangent space $\xi ^1,\xi ^2,...,\xi ^n$, running
through all the points of the indicatrix, then in the Finsler space  {%
it} is possible (similar to (\ref{10})) to define the volume element $dV$ by
\begin{equation}
dV=const\cdot \frac{dx^1dx^2...dx^n}{\left( V_{ind}\right) _{{\small {eu}}}}%
\,,  \label{11}
\end{equation}
where $\left( V_{ind}\right) _{{eu}}$ is the volume of  {the} indicatrix, calculated
under the assumption, that the tangent space is Euclidean and the coordinates are
Cartesian. This is quite evident, that volume element, defined in this way is
invariant with respect to coordinate transformations.

Consider $n$-dimensional Riemannian space. In this case the metric function is given
by
\begin{equation}
L(dx;x)=\sqrt{g_{ij}dx^idx^j}\,,  \label{12}
\end{equation}
and the equation of the indicatrix is given by
\begin{equation}
g_{ij}\xi ^i\xi ^j=1\,.  \label{13}
\end{equation}
This equation defines the hypersurface of order 2, namely  {the }%
ellipsoid. If the space $\xi ^1,\xi ^2,...,\xi ^n$ is Euclidean, then the
volume of this ellipsoid is equal to
\begin{equation}
\left( V_{ind}\right) _{{\small {eu}}}=\frac{const^{\prime }}{\sqrt{%
det(g_{ij})}}\,.  \label{14}
\end{equation}
Substituting the last relation into (\ref{11}), we obtain the formula for
the volume element in an arbitrary Riemannian space:
\begin{equation}
dV=const\cdot \sqrt{det(g_{ij})}\quad dx^1dx^2...dx^n\,,  \label{15}
\end{equation}
This relation is a conventional definition of invariant volume element in the
Riemannian space.

For pseudo-Riemannian spaces, when there are no additional constraints on the
indicatrix, we get
\begin{equation}
\left( V_{ind}\right) _{{\small {eu}}}=\infty \qquad \Rightarrow \qquad
dV=0\cdot dx^{1}dx^{2}...dx^{n}\,.  \label{16}
\end{equation}
But this is possible to provide the line of reasoning which allows one to propose
for pseudo-Riemannian space the definition of the invariant volume element in the
form, similar to(\ref{15}). The same reasoning should be provided to obtain the
invariant volume element in Finsler spaces, where the problem (\ref{16}) takes
place. As a start, we should consider some flat space, close to the space, where the
volume element should be defined.

We will provide this reasoning for a particular example: for pseudo-Riemannian space
with the signature $(+,-,-,-)$. In this case we will start with Minkowski space
$x^0,x^1,x^2,x^3$, with the metric function of the form
\begin{equation}
L(dx)=\sqrt{(dx^0)^2-(dx^1)^2-(dx^2)^2-(dx^3)^2}\equiv \sqrt{\stackrel{o}{g}%
_{ij}dx^idx^j}\,,  \label{17}
\end{equation}
and with the tangential equation of the indicatrix in the form
\begin{equation}
(\xi ^0)^2-(\xi ^1)^2-(\xi ^2)^2-(\xi ^3)^2=1.  \label{18}
\end{equation}
This equation  {is }of the second order and  {it }defines the hypersurface, which is
a hyperboloid of two sheets; thus, the problem of calculating the volume of the
indicatrix does take place. As both the metric function and indicatrix equation are
the same for all the point in the space, then regardless of how the corresponding
integral is regularized, we will obtain a real number, the same for all the points
in the space . Let  {us }denote this number by $\left( V_{ind}\right) _{{eu}}$. In
order to obtain the invariant volume element  in Minkowski space using (\ref{11}),
the quantity $\left( V_{ind}\right) _{{eu}}$ should be written in the form
\begin{equation}
\left( V_{ind}\right) _{{\small {eu}}}=\frac{const^{\prime }}{\sqrt{%
-det\left( \stackrel{o}{g}_{ij}\right) }}\,.  \label{19}
\end{equation}
Now we change the coordinates \ $x^0,x^1,x^2,x^3$ \thinspace \ to curvilinear
coordinates \ $x^{0^{\prime }},x^{1^{\prime }},x^{2^{\prime }},x^{3^{\prime }}$. As
a result, $\stackrel{o}{g}_{ij}$ will be substituted by $g(x^{\prime })_{i^{\prime
}j^{\prime }}$, and  {the }volume element in Minkowski space in the curvilinear
coordinates $x^{0^{\prime }},\,x^{1^{\prime }},\,x^{2^{\prime }},\,x^{3^{\prime }}$
will be given by
\begin{equation}
dV=const\cdot \sqrt{-det\left( g(x^{\prime })_{i^{\prime }j^{\prime
}}\right) }\quad dx^{0^{\prime }}dx^{1^{\prime }}dx^{2^{\prime
}}dx^{3^{\prime }}\,,  \label{20}
\end{equation}
but this is still the same Minkowski space.

Consider the pseudo-Riemannian space which is conformally connected \cite{4} with
Minkowski space
\begin{equation}
ds=\kappa (x)\cdot \sqrt{\stackrel{o}{g}_{ij}dx^idx^j}\,,  \label{21}
\end{equation}
where $\kappa (x)>0$. This space cannot be converted to the Minkowski space by any
coordinate transform.  The indicatrix equation for this pseudo-Riemannian space may
be written in the form:
\begin{equation}
(\xi ^0)^2-(\xi ^1)^2-(\xi ^2)^2-(\xi ^3)^2=\frac 1{\kappa ^2(x)}\,.
\label{22}
\end{equation}
 {Comparing} (\ref{22}) with the equation (\ref{18}), one  {can}
notice, that the hypersurface, given by (\ref{22}),  {can} be obtained from the
hypersurface given by (\ref{18}), via scaling coordinate transform with the
coefficient $\frac 1{\kappa (x)}$. Thus, if we assign volume (\ref {19}) to the
indicatrix (\ref{18}), then to the indicatrix (\ref{22}) we should assign the volume
by
\begin{equation}
\left( V_{ind}\right) _{{\small {eu}}}=\frac{const^{\prime }}{\kappa ^4(x)%
\sqrt{-det\left( \stackrel{o}{g}_{ij}\right) }}=\frac{const^{\prime }}{\sqrt{%
-det\left( g(x)_{ij}\right) }}\,,  \label{23}
\end{equation}
where
\begin{equation}
g(x)_{ij}\equiv \kappa ^2(x)\stackrel{o}{g}_{ij}\,.  \label{24}
\end{equation}

From the reasoning provided above, it follows that in the pseudo-Riemannian space
with the metric tensor $g(x)_{ij}$ and the signature $(+,-,-,-)$ this is possible to
define the volume element by
\begin{equation}
dV=const\cdot \sqrt{-det\left( g(x)_{ij}\right) }\quad dx^0dx^1dx^2dx^3\,,
\label{25}
\end{equation}
 {and this }corresponds to the approach conventional for GRT\cite{1} .

The problem (\ref{16}) in pseudo-Riemannian spaces may be handled more rigorously
(however, this is outside the scope of this paper), but we will have to deal with
the spaces  {that} are more general than pseudo-Riemannian spaces. This may be
explained on the example of Minkowski space. If instead of Minkowski space with the
metric function (\ref{17}) we consider the Finsler space with the metric function
\begin{equation}
L(dx)=\sqrt{(dx^0)^2-(dx^1)^2-(dx^2)^2-(dx^3)^2}+q_0dx^0  \label{26}
\end{equation}
and the constraint $dx^0\geq 0$, where $q_0>0$, then for this space the volume of
the indicatrix $\left( V_{ind}\right) _{{eu}}$ will be a finite real number,
depending on the parameter $q_0,$ such that $\left( V_{ind}\right) _{{eu}}$tends to
$\infty $, as the parameter $q_0$ vanishes.

Thus, we will assume that in any Finsler space, where the problem (%
\ref{16}) takes place, this problem is solvable. Then, this is possible to
claim that if the metric function of this space contains certain fields, the
geometry of Finsler space yields automatically the Lagrangian
\begin{equation}
\mathfrak{L}=\frac{const}{\left( V_{ind}\right) _{{\small {eu}}}}\,,
\label{27}
\end{equation}
and from this Lagrangian one can obtain the field equations.

  \medskip
\textit{Remark}. Henceforth the constants which appear in the relations (\ref{11}),
(\ref{14}),..., (\ref{27}) will be omitted, as these constants are not involved in
the the field equations.

\section{The spaces, conformally connected to Euclidean spaces}

In the space conformally connected to $n$-dimensional Euclidean space, the length
element is given by
\begin{equation}
ds=\kappa (x)\cdot \sqrt{(x^{1})^{2}+(x^{2})^{2}+...+(x^{n})^{2}}\,,
\label{28}
\end{equation}
where $\kappa (x)>0$. As in this case the following relation holds,
\begin{equation}
\sqrt{det(g_{ij})}=\kappa ^{n}(x)\,,  \label{29}
\end{equation}
the Lagrangian takes the form
\begin{equation}
\mathfrak{L}=\kappa ^{n}(x)\,.  \label{30}
\end{equation}

To construct the field equation  {with the help of} this Lagrangian,  {it} is
necessary to represent the scalar field $\kappa (x)$ in terms of another field so
that the lagrangian will involve the derivatives of the new field. A method to
achieve this goal is proposed in \cite{5}, \cite{6}.

The generalized momenta in the space (\ref{28}) are given by
\begin{equation}
p_{i}=\kappa (x)\frac{dx^{i}}{\sqrt{(x^{1})^{2}+(x^{2})^{2}+...+(x^{n})^{2}}}%
\,,  \label{31}
\end{equation}
and the tangential equation of the indicatrix may be written in the form:
\begin{equation}
p_{1}^{2}+p_{2}^{2}+...+p_{n}^{2}-\kappa ^{2}(x)=0\,.  \label{32}
\end{equation}
Consider scalar function $S(x)$, which in the space $x^{1},x^{2},...,x^{n}$.
Let this function define the normal congruence of geodesics: in classical
mechanics this function is called the `action as a function of coordinates',
and in the paper \cite{5} the function $S(x)$ is called the World function.
This function must satisfy the Hamilton-Jacobi equation
\begin{equation}
\left( \frac{\partial S}{\partial x^{1}}\right) ^{2}+\left( \frac{\partial S%
}{\partial x^{2}}\right) ^{2}+...+\left( \frac{\partial S}{\partial x^{n}}%
\right) ^{2}=\kappa ^{2}(x)\,.  \label{33}
\end{equation}
Thus,
\begin{equation}
\mathfrak{L}=\left[ \left( \frac{\partial S}{\partial x^{1}}\right)
^{2}+\left( \frac{\partial S}{\partial x^{2}}\right) ^{2}+...+\left( \frac{%
\partial S}{\partial x^{n}}\right) ^{2}\right] ^{\frac{n}{2}}\,,  \label{34}
\end{equation}
and the field equation (\ref{4}) takes the form:
\begin{equation}
\frac{\partial }{\partial x^{i}}\left\{ \frac{\partial S}{\partial x^{i}}%
\left[ \left( \frac{\partial S}{\partial x^{1}}\right) ^{2}+\left( \frac{%
\partial S}{\partial x^{2}}\right) ^{2}+...+\left( \frac{\partial S}{%
\partial x^{n}}\right) ^{2}\right] ^{\frac{n}{2}-1}\right\} =0\,.  \label{35}
\end{equation}
Note, that for $n>2$ this equation is a non-linear partial differential equation of
second order.

For the space, conformally connected to the 2-dimensional Euclidean plain $(x,y)$,
the equation (\ref{35}) may be written in the form
\begin{equation}
\frac{\partial ^{2}S}{\partial x^{2}}+\frac{\partial ^{2}S}{\partial y^{2}}%
=0\,,  \label{36}
\end{equation}
i.e. the function $S(x,y)$ satisfies the Laplace equation; therefore this
function is a component of the analytical function of complex variable.
Thus,
\begin{equation}
\kappa (x,y)=\sqrt{\left( \frac{\partial S}{\partial x}\right) ^{2}+\left(
\frac{\partial S}{\partial y}\right) ^{2}}  \label{37}
\end{equation}
is the factor of the conformal transformation of the length element in the
Euclidean space
\begin{equation}
ds^{\prime }=\sqrt{(x^{\prime 2}+(y^{\prime 2}}=\kappa (x,y)\sqrt{x^{2}+y^{2}%
}  \label{38}
\end{equation}
for the conformal transformation
\begin{equation}
x^{\prime }=u(x,y)\,,\qquad y^{\prime }=\pm v(x,y)\,,  \label{39}
\end{equation}
where the function $S$ is one of the components of the analytical function $%
u+iv$ of complex variable $x+iy$.

\medskip

Now, we will solve the equation (\ref{35}) under the assumption that function $S$ is
a function of radius only
\begin{equation}
r=\sqrt{(x^1)^2+(x^2)^2+...+(x^n)^2}\,.  \label{40}
\end{equation}
To find the solution, this will be more convenient, if volume element is
represented as a function of spherical coordinates. Then, after integration
on all the angles, we obtain
\begin{equation}
dV_r=r^{n-1}\left| \frac{dS}{dr}\right| ^n\,dr\qquad \Rightarrow \qquad %
\mathfrak{L}_r=r^{n-1}\left| \frac{dS}{dr}\right| ^n\,.  \label{41}
\end{equation}
Then the field equation will take the form :
\begin{equation}
\frac d{dr}\left[ r^{n-1}\left| \frac{dS}{dr}\right| ^{n-1}\right] =0\,.
\label{42}
\end{equation}
Via integration of the last relation, we get
\begin{equation}
\frac{dS}{dr}=\frac Cr\,,\qquad S=C\ln \frac r{r_0}\,,  \label{43}
\end{equation}
where $C\neq 0$, $r_0>0$ are real. Thus,
\begin{equation}
\kappa (x)=\left| \frac{dS}{dr}\right| =\frac{|C|}r\,.  \label{44}
\end{equation}

In this space the geodesics are given by the relations
\begin{equation}
\dot{x}^{i}=\frac{dS}{dx^{i}}\cdot \lambda (x)\,,  \label{45}
\end{equation}
where $\lambda (x)\neq 0$ is a function, $\dot{x}^{i}$ is the parameter
derivative of $x^{i}$ along the geodesic $\tau $. Set $\lambda (x)=r$, then
the relation (\ref{45}) yields
\begin{equation}
\dot{x}^{i}=x^{i}\,.  \label{46}
\end{equation}
Let $j>1$, then
\begin{equation}
\frac{dx^{j}}{dx^{1}}=\frac{x^{j}}{x^{1}}\qquad \Rightarrow \qquad
x^{j}=C^{j}x^{1}\,,  \label{47}
\end{equation}
thus, the geodesics in this space are straight lines, going through the origin with
the directing vector $(1,C^{2},C^{3},...,C^{n})$.

\section{The spaces, conformally connected to
pseudo-Euclidean spaces with the signature $(+,-,-,...,-)$}

In the space, which is conformally connected to the $n$%
-dimensional pseudo-Euclidean space with the signature $(+,-,-,...,-)$ the
length element is given by
\begin{equation}
ds=\kappa (x)\cdot \sqrt{(x^{0})^{2}-(x^{1})^{2}-...-(x^{n-1})^{2}}\,,
\label{48}
\end{equation}
where $\kappa (x)>0$. As in this case the following relation holds
\begin{equation}
\sqrt{(-1)^{n-1}det\left( g_{ij}\right) }=\kappa ^{n}(x)\,,  \label{49}
\end{equation}
the Lagrangian can be represented in the form
\begin{equation}
\mathfrak{L}=\kappa ^{n}(x)\,.  \label{50}
\end{equation}

In order to construct the field equation from this Lagrangian,  {it} is required to
express the scalar field $\kappa (x)$ via another field so that, the Lagrangian will
contain the derivatives of the new field \cite{5}, \cite{6}.

The generalized momenta in the space (\ref{48}) are given by:
\begin{equation}
p_{0}=\frac{\kappa (x)\;dx^{0}}{\sqrt{%
(x^{0})^{2}-(x^{1})^{2}-...-(x^{n-1})^{2}}},\,p_{\mu }=-\frac{\kappa
(x)\;dx^{\mu }}{\sqrt{(x^{0})^{2}-(x^{1})^{2}-...-(x^{n-1})^{2}}},
\label{51}
\end{equation}
where $\mu =1,2,...,(n-1)$, and tangential equation of the indicatrix may be
represented in the form:
\begin{equation}
p_{0}^{2}-p_{1}^{2}-...-p_{n-1}^{2}-\kappa ^{2}(x)=0\,.  \label{52}
\end{equation}
The scalar function $S(x)$, which in the space $x^{0},x^{1},...,x^{n-1}$
defines the normal congruence of geodesics, must satisfy Hamilton-Jacobi
equation
\begin{equation}
\left( \frac{\partial S}{\partial x^{0}}\right) ^{2}-\left( \frac{\partial S%
}{\partial x^{1}}\right) ^{2}-...-\left( \frac{\partial S}{\partial x^{n-1}}%
\right) ^{2}=\kappa ^{2}(x)\,.  \label{53}
\end{equation}
Thus,
\begin{equation}
\mathfrak{L}=\left[ \left( \frac{\partial S}{\partial x^{0}}\right)
^{2}-\left( \frac{\partial S}{\partial x^{1}}\right) ^{2}-...-\left( \frac{%
\partial S}{\partial x^{n-1}}\right) ^{2}\right] ^{\frac{n}{2}}\,,
\label{54}
\end{equation}
and the field equation (\ref{4}) takes the form:
\begin{equation}
\begin{array}{l}
\quad \displaystyle\frac{\partial }{\partial x^{0}}\left\{ \frac{\partial S}{%
\partial x^{0}}\left[ \left( \frac{\partial S}{\partial x^{0}}\right)
^{2}-\left( \frac{\partial S}{\partial x^{1}}\right) ^{2}-...-\left( \frac{%
\partial S}{\partial x^{n-1}}\right) ^{2}\right] ^{\frac{n}{2}-1}\right\} -
\\[24pt]
-\displaystyle\frac{\partial }{\partial x^{\mu }}\left\{ \frac{\partial S}{%
\partial x^{\mu }}\left[ \left( \frac{\partial S}{\partial x^{0}}\right)
^{1}-\left( \frac{\partial S}{\partial x^{2}}\right) ^{2}-...-\left( \frac{%
\partial S}{\partial x^{n-1}}\right) ^{2}\right] ^{\frac{n}{2}-1}\right\}
=0\,.
\end{array}
\label{55}
\end{equation}
Interestingly, that for $n>2$ this equation is a non-linear partial
differential equations of second order and this equation is satisfied if the
function $S$ satisfies the eikonal equation
\[
\left( \frac{\partial S}{\partial x^{0}}\right) ^{1}-\left( \frac{\partial S%
}{\partial x^{2}}\right) ^{2}-...-\left( \frac{\partial S}{\partial x^{n-1}}%
\right) ^{2}=0\,.
\]
For the field equation (\ref{55}) to be the wave equation, the function $S$
must simultaneously satisfy one more condition:
\[
\left( \frac{\partial S}{\partial x^{0}}\right) ^{1}-\left( \frac{\partial S%
}{\partial x^{2}}\right) ^{2}-...-\left( \frac{\partial S}{\partial x^{n-1}}%
\right) ^{2}=const\,.
\]

For the space conformally connected with the 2-dimensional pseudo-Euclidean plain
$(x,y)$, the relation (\ref{55}) takes the form
\begin{equation}
\frac{\partial ^{2}S}{\partial x^{2}}-\frac{\partial ^{2}S}{\partial y^{2}}%
=0\,,  \label{56}
\end{equation}
that is for the two-dimensional case the field equation (\ref{55}) is a wave
equation.

\medskip

Now we will solve the equation (\ref{55}) under the assumption that the function $S$
depends only on the interval

\begin{equation}
s=\sqrt{(x^{0})^{2}-(x^{1})^{2}-...-(x^{n-1})^{2}}\,.  \label{57}
\end{equation}
For this we will consider the volume element, choosing as one of the
variables the interval $s$. In the process of integration on hyperbolic
angles certain difficulties may take place, which are similar to (\ref{16})
and which can be resolved in the similar way, thus
\begin{equation}
dV_{s}=s^{n-1}\left\vert \frac{dS}{ds}\right\vert ^{n}\,ds\qquad \Rightarrow
\qquad \mathfrak{L}_{s}=s^{n-1}\left\vert \frac{dS}{ds}\right\vert ^{n}\,,
\label{58}
\end{equation}
and the field equation takes the form:
\begin{equation}
\frac{d}{ds}\left[ s^{n-1}\left\vert \frac{dS}{ds}\right\vert ^{n-1}\right]
=0\,.  \label{59}
\end{equation}
Integrating the last equality, we obtain
\begin{equation}
\frac{dS}{ds}=\frac{C}{s}\,,\qquad S=C\ln \frac{s}{s_{0}}\,,  \label{60}
\end{equation}
where $C\neq 0$, $s_{0}>0$ are real. Thus,
\begin{equation}
\kappa (x)=\left\vert \frac{dS}{ds}\right\vert =\frac{|C|}{s}\,.  \label{61}
\end{equation}

The geodesics in this space are given by
\begin{equation}
\dot{x}^{0}=\frac{dS}{dx^{0}}\cdot \lambda (x)\,,\qquad \dot{x}^{\mu }=-%
\frac{dS}{dx^{\mu }}\cdot \lambda (x)\,,  \label{62}
\end{equation}
where $\lambda (x)\neq 0$ is a function , $\dot{x}^{i}$ is a derivative of $%
x^{i}$ with respect to the evolution parameter $\tau $, $\mu =1,2,...,n-1$.
Set $\lambda (x)=\displaystyle\frac{s^{2}}{|C|}$, then from (\ref{62}) it
follows that
\begin{equation}
\dot{x}^{i}=x^{i}\,,  \label{63}
\end{equation}
or
\begin{equation}
\frac{dx^{\mu }}{dx^{0}}=\frac{x^{\mu }}{x^{0}}\qquad \Rightarrow \qquad
x^{\mu }=C^{\mu }x^{0}\,,  \label{64}
\end{equation}
that is the geodesics (extremals) in this space are straight lines, `going' through
the origin with the directing vector $(1,C^{2},C^{3},...,C^{n})$. The interval will
also change linearly with respect to the coordinate $x^{0}$,
\begin{equation}
s=\sqrt{1-(C^{1})^{2}-...-(C^{n-1})^{2}}\cdot x^{0}\,,  \label{65}
\end{equation}
$x^{0}>0$.

\medskip

 As we will further use the space, which is conformally connected to the Minkowski
space, for construction of the cosmological equation, we will provide certain
formulae of this section for $n=4$, using the metric tensor of Minkowski space
$\stackrel{o}{g}_{ij}$:

Relation between the function $S(x)$ and the factor $\kappa (x)$:
\begin{equation}
\stackrel{o}{g}^{\;ij}\frac{\partial S}{\partial x^{i}}\frac{\partial S}{%
\partial x^{j}}=\kappa ^{2}(x)\,,  \label{66}
\end{equation}

 Lagrangian:
\begin{equation}
\mathfrak{L}=\left( \stackrel{o}{g}^{\;ij}\frac{\partial S}{\partial x^{i}}%
\frac{\partial S}{\partial x^{j}}\right) ^{2}\,,  \label{67}
\end{equation}

 Field equation:
\begin{equation}
\stackrel{o}{g}^{\;kl}\displaystyle\frac{\partial }{\partial x^{k}}\left[
\frac{\partial S}{\partial x^{l}}\left( \stackrel{o}{g}^{\;ij}\frac{\partial
S}{\partial x^{i}}\frac{\partial S}{\partial x^{j}}\right) \right] =0\,.
\label{68}
\end{equation}

\section{Model cosmological equation in the space,\\
conformally connected to the Minkowski space}

We will write the equation (\ref{68}) under the assumption that the function $S$ is
of the form
\begin{equation}
S(x^{0},r)=S_{0}e^{-\gamma x^{0}}\psi (r)\,,  \label{69}
\end{equation}
where $r=\sqrt{(x^{1})^{2}+(x^{2})^{2}+(x^{3})^{2}}$, and $\gamma $ , $S_{0}$%
\ \ are constant. This is much simpler to obtain the field equation of this
form, if in relations for volume element instead of spatial coordinates $%
x^{1},x^{2},x^{3}$ the spherical coordinate system is used. After
integration on spherical angles (omitting the constant), we will obtain the
following relation for the Lagrangian
\begin{equation}
\mathfrak{L}=r^{2}\left[ \left( \frac{\partial S}{\partial x^{0}}\right)
^{2}-\left( \frac{\partial S}{\partial r}\right) ^{2}\right] ^{2}\,,
\label{70}
\end{equation}
and the field equation will take the form:
\begin{equation}
r^{2}\frac{\partial }{\partial x^{0}}\left\{ \frac{\partial S}{\partial x^{0}%
}\left[ \left( \frac{\partial S}{\partial x^{0}}\right) ^{2}-\left( \frac{%
\partial S}{\partial r}\right) ^{2}\right] \right\} -\frac{\partial }{%
\partial r}\left\{ r^{2}\frac{\partial S}{\partial r}\left[ \left( \frac{%
\partial S}{\partial x^{0}}\right) ^{2}-\left( \frac{\partial S}{\partial r}%
\right) ^{2}\right] \right\} =0\,.  \label{71}
\end{equation}
Substituting into the last relation the function $S(x^{0},r)$ (\ref{69}), we
obtain
\begin{equation}
3\gamma ^{2}r^{2}\psi \left[ \gamma ^{2}\psi ^{2}-\left( \frac{d\psi }{dr}%
\right) ^{2}\right] -\frac{d}{dr}\left\{ r^{2}\frac{d\psi }{dr}\left[ \gamma
^{2}\psi ^{2}-\left( \frac{d\psi }{dr}\right) ^{2}\right] \right\} =0\,.
\label{72}
\end{equation}
Let define the dimensionless variable $\xi \equiv \gamma r$, then the last
equation may be rewritten in the from:
\begin{equation}
3\xi ^{2}\psi \left[ \psi ^{2}-\left( \frac{d\psi }{d\xi }\right)
^{2}\right] -\frac{d}{d\xi }\left\{ \xi ^{2}\frac{d\psi }{d\xi }\left[ \psi
^{2}-\left( \frac{d\psi }{d\xi }\right) ^{2}\right] \right\} =0\,.
\label{73}
\end{equation}
As this equation is homogeneous with respect to the unknown function$\psi
(\xi )$, we will suppose the solution to be of the form
\begin{equation}
\psi (\xi )=\psi _{0}\exp \left( \int\limits_{0}^{\xi }\varphi (\xi )d\xi
\right) \,,  \label{74}
\end{equation}
where $\psi _{0}$ is a constant. This constant for construction of the
function $S$ is multiplied by the constant $S_{0}$, thus we will set $\psi
_{0}=1$. Substituting (\ref{74}) into (\ref{73}), we obtain
\begin{equation}
\frac{d}{d\xi }\left[ \xi ^{2}\varphi (1-\varphi ^{2})\right] -3\xi
^{2}(1-\varphi ^{2})^{2}=0\,,  \label{75}
\end{equation}
or
\begin{equation}
\xi (1-3\varphi ^{2})\frac{d\varphi }{d\xi }+2\varphi (1-\varphi ^{2})-3\xi
(1-\varphi ^{2})^{2}=0\,.  \label{76}
\end{equation}
There were no success in finding the analytical solution of the last equation.

For the domain $\xi \ll 1$ we will find a solution in a form of power series
\begin{equation}
\varphi \simeq A\xi +B\xi ^{2}+C\xi ^{3}+O(\xi ^{4})\,.  \label{77}
\end{equation}
Substituting this expansion into equation (\ref{76}), after grouping the
terms, we will obtain
\begin{equation}
\varphi \simeq \xi -\frac{1}{5}\xi ^{3}+O(\xi ^{4})\,.  \label{78}
\end{equation}

The sample bodies (stars) move along the geodesics (extremals) of the space with the
length element
\begin{equation}
ds=\kappa (x^{0},r)\sqrt{(dx^{0})^{2}-(dr)^{2}}  \label{79}
\end{equation}
and the tangential equation of indicatrix
\begin{equation}
p_{0}^{2}-p_{r}^{2}=\kappa ^{2}(x^{0},r)\,.  \label{80}
\end{equation}
For the field $S$ (\ref{69}), (\ref{74}) the scaling factor of conformal
transformation may be calculated as
\begin{equation}
\kappa (x^{0},r)=\sqrt{\left( \frac{\partial S}{\partial x^{0}}\right)
^{2}-\left( \frac{\partial S}{\partial r}\right) ^{2}}=\gamma \cdot \sqrt{%
1-\varphi ^{2}}\cdot S(x^{0},r)\,.  \label{81}
\end{equation}
From the last relation, it follows that $|\varphi |<1$. The motion equations in this
case will be of the form
\begin{equation}
\dot{x}^{0}=\frac{\partial S}{\partial x^{0}}\lambda =-\gamma S\lambda
\,,\qquad \dot{r}=-\frac{\partial S}{\partial r}\lambda =-\gamma S\varphi
(\gamma r)\lambda \,,  \label{82}
\end{equation}
where the dot represents the total derivative with respect to certain
evolution parameter $\tau $, and an arbitrary function $\lambda \neq 0$.
Then
\begin{equation}
\frac{dr}{dx^{0}}=\varphi (\gamma r)\qquad \Rightarrow \qquad \frac{dr}{dt}%
=c\varphi (\gamma r)\,.  \label{83}
\end{equation}
As $|\varphi |<1$, then
\[
\left\vert \frac{dr}{dx^{0}}\right\vert <1\qquad \hbox{and}\qquad \left\vert
\frac{dr}{dt}\right\vert <c\,.
\]
Let consider, the behavior of the velocity of the sample body in the domain $%
\xi \ll 1$, for this we substitute (\ref{78}) into the obtained relation:
\begin{equation}
\frac{dr}{dt}=c\gamma \left( 1-\frac{1}{5}\gamma ^{2}r^{2}\right) \cdot r\,.
\label{84}
\end{equation}
If we denote the Hubble's constant by $H_{0}$, then according to the
obtained relation the Hubble law holds when $\gamma r<\frac{1}{10}$, and $%
H_{0}=c\gamma $, and the tendence, how the `Hubble constant' $H(r)$ evolves
initially as the distance from the center grows is of the form:
\begin{equation}
\frac{dr}{dt}=H(r)\cdot r\,,\qquad H(r)=H_{0}\cdot \left[ 1-\frac{1}{5}%
\left( \frac{H_{0}}{c}\right) ^{2}\cdot r^{2}\right] \,.  \label{85}
\end{equation}
I.e. in the domain $\xi \ll 1$ this constant $H(r)$ decreases as the distance from
the origin grows.

To provide any ideas about  {the }size of the universe and the dependence $H(r)$ for
all possible values of the variable $r$, this is
required to analyze the solution $\varphi (\xi )$ of the equation (\ref{76}%
), the solution which (as $\xi \rightarrow 0)$ takes the form (\ref{78}).
Neither analytically, nor numerically we didnot succeed in this analysis, as
approaching the value $\varphi =\frac 1{\sqrt{3}}$ the behavior of the
solution becomes quite complicated (unstable). If we suppose the the
solution of the equation (\ref{76}) can be obtained and analyzed, then
general form of the quantity $H(r)$ may be written in the following way:
\begin{equation}
H(r)=H_0\cdot \left[ \frac{\varphi \left( \frac{H_0}cr\right) }{\frac{H_0}cr}%
\right] \,.  \label{86}
\end{equation}

If we consider motion trajectories in the space $%
x^{0},x^{1},x^{2},x^{3}$ with the World function $S$ (\ref{69}), these
trajectories will be given by the equations
\[
\frac{dx^{\mu }}{dx^{0}}=\varphi (\gamma r)\,\frac{x^{\mu }}{r}\,,
\]
that is the motion will be along the rays from the origin, and this means that the
sample particle move rectilinearly, but certainly the motion will be still
non-uniform.

\medskip

As the space with the length element (\ref{79}) is a pseudo-Riemannian space with
the metric tensor
\begin{equation}
g_{ij}(x^{0},r)=\kappa ^{2}(x^{0},r)\cdot \stackrel{o}{g}_{ij}\,,  \label{87}
\end{equation}
where $\stackrel{o}{g}_{ij}$ is the metric tensor in the Minkowski space and
\begin{equation}
\kappa (x^{0},r)=\gamma S\sqrt{1-\varphi ^{2}}\,,  \label{88}
\end{equation}
then for this space this is possible to calculate the curvature tensor and its
contractions, and directly from the Einstein equations one may obtain the matter
energy-momentum tensor $T_{km}$, which is involved in the Einstein equations and
which corresponds to the space with the metric tensor (\ref{87}). Interestingly,
that the equations for the gravitational field, certainly, for this energy-momentum
tensor will hold automatically, but with the tensor $T_{km}$ this is not possible,
in general, to associate the laws of conservation of energy and momentum.

Let  {us }introduce a new quantity, which can be employed quite usefully
\begin{equation}
a=\ln (\kappa ^2/const)\,.  \label{89}
\end{equation}
Then, using the well-known classical formulae, we obtain the expressions for
the connectivity object:
\begin{equation}
\Gamma _{kl}^i=\frac 12\left( \frac{\partial a}{\partial x^l}\delta _k^i+%
\frac{\partial a}{\partial x^k}\delta _l^i-\stackrel{o}{g}^{is}\frac{%
\partial a}{\partial x^s}\stackrel{o}{g}_{kl}\right) \,,  \label{90}
\end{equation}
curvature tensor:
\begin{equation}
\begin{array}{l}
R_{klm}^i= \\[12pt]
=\displaystyle\frac 12\left( \frac{\partial ^2a}{\partial x^l\partial x^k}%
\,\delta _m^i-\frac{\partial ^2a}{\partial x^k\partial x^m}\,\delta _l^i-%
\stackrel{o}{g}^{\,is}\frac{\partial ^2a}{\partial x^l\partial x^s}\stackrel{%
o}{g}_{km}+\stackrel{o}{g}^{\,is}\frac{\partial ^2a}{\partial x^m\partial x^s%
}\stackrel{o}{g}_{kl}\right) + \\[16pt]
\displaystyle\frac 14\left( \frac{\partial a}{\partial x^m}\frac{\partial a}{%
\partial x^k}\,\delta _l^i-\frac{\partial a}{\partial x^l}\frac{\partial a}{%
\partial x^k}\,\delta _m^i-\stackrel{o}{g}^{\,ns}\frac{\partial a}{\partial
x^n}\frac{\partial a}{\partial x^s}\,\delta _l^i\stackrel{o}{g}_{km}+\right.
 \\[16pt]
\displaystyle\left. +\frac{\partial a}{\partial x^l}\stackrel{o}{g}_{km}%
\stackrel{o}{g}^{\,is}\frac{\partial a}{\partial x^s}+\stackrel{o}{g}^{\,ns}%
\frac{\partial a}{\partial x^n}\frac{\partial a}{\partial x^s}\,\delta _m^i%
\stackrel{o}{g}_{kl}-\frac{\partial a}{\partial x^m}\stackrel{o}{g}_{kl}%
\stackrel{o}{g}^{\,is}\frac{\partial a}{\partial x^s}\right) \,,
\end{array}
\label{91}
\end{equation}
Ricci tensor:
\begin{equation}
\begin{array}{l}
R_{km}\equiv R_{klm}^l= \\[12pt]
=\displaystyle\frac 12\left( -2\frac{\partial ^2a}{\partial x^k\partial x^m}-%
\stackrel{o}{g}^{\,ns}\frac{\partial ^2a}{\partial x^n\partial x^s}\stackrel{%
o}{g}_{km}+\frac{\partial a}{\partial x^k}\frac{\partial a}{\partial x^m}-%
\stackrel{o}{g}^{\,ns}\frac{\partial a}{\partial x^n}\frac{\partial a}{%
\partial x^s}\stackrel{o}{g}_{km}\right) \,,
\end{array}
\label{92}
\end{equation}
scalar curvature of the space:
\begin{equation}
R\equiv g^{km}R_{km}=\frac 1{\kappa ^2}\stackrel{o}{g}^{km}R_{km}=-\frac
3{\kappa ^2}\left( 2\stackrel{o}{g}^{km}\frac{\partial ^2a}{\partial
x^k\partial x^m}+\stackrel{o}{g}^{km}\frac{\partial a}{\partial x^k}\frac{%
\partial a}{\partial x^m}\right) \,,  \label{93}
\end{equation}
matter energy-momentum tensor:
\begin{equation}
T_{km}=\frac{c^4}{8\pi k}\left( R_{km}-\frac 12\kappa ^2\stackrel{o}{g}%
_{km}R\right) \,,  \label{94}
\end{equation}
where $k$ is the gravitation constant. Hence,
\begin{equation}
T\equiv g^{km}T_{km}=\frac 1{\kappa ^2}\stackrel{o}{g}^{km}T_{km}=-\frac{c^4%
}{8\pi k}R\,.  \label{95}
\end{equation}

But using the `independence' on the Einstain gravitation field equations, we can
calculate the full energy-momentum tensor $\hat{T}_{km}$. For the Lagrangian of the
field $\mathfrak{L}$ (\ref{67}) we obtain
\begin{equation}
\hat{T}_{m}^{k}=\frac{\partial S}{\partial x^{m}}\frac{\partial \mathfrak{L}%
}{\partial \frac{\partial S}{\partial x^{k}}}-\delta _{m}^{k}\mathfrak{L}=4%
\stackrel{o}{g}^{ks}\frac{\partial S}{\partial x^{s}}\frac{\partial S}{%
\partial x^{m}}\left( \stackrel{o}{g}^{rs}\frac{\partial S}{\partial x^{r}}%
\frac{\partial S}{\partial x^{s}}\right) -\delta _{m}^{k}\left( \stackrel{o}{%
g}^{rs}\frac{\partial S}{\partial x^{r}}\frac{\partial S}{\partial x^{s}}%
\right) ^{2}\,,  \label{96}
\end{equation}
after contraction on 2 used indices, we get
\begin{equation}
\hat{T}_{k}^{k}\equiv 0\,.  \label{97}
\end{equation}

\medskip

Finally, one may note that the tensors $T_{km}$ and $\hat{T}_{km}$ are essentially
different.

\section{The space, conformally connected\\ to 4-dimensional
Berwald-Moore space}

The length element in this space (in the special isotropic basis) will have the form
\begin{equation}
ds=\kappa (\xi ^{1},\xi ^{2},\xi ^{3},\xi ^{4})\sqrt[4]{d\xi ^{1}d\xi
^{2}d\xi ^{3}d\xi ^{4}}\,.  \label{98}
\end{equation}
The generalized momenta will satisfy the relations
\begin{equation}
p_{i}=\frac{1}{4}\kappa (\xi )\frac{\sqrt[4]{d\xi ^{1}d\xi ^{2}d\xi ^{3}d\xi
^{4}}}{d\xi ^{i}}\,.  \label{99}
\end{equation}

If $\eta ^{1},\eta ^{2},\eta ^{3},\eta ^{4}$ are coordinates of tangent centroaffine
space in the point $M(\xi ^{1},\xi ^{2},\xi ^{3},\xi ^{4})$ of the main space, then
the indicatrix equation will have the form
\begin{equation}
\eta ^{1}\eta ^{2}\eta ^{3}\eta ^{4}=\frac{1}{\kappa ^{4}(\xi )}\,,
\label{100}
\end{equation}
and the tangential equation of indicatrix will have e.g. the form,
\begin{equation}
p_{1}p_{2}p_{3}p_{4}=\frac{\kappa ^{4}(\xi )}{4^{4}}\,.  \label{101}
\end{equation}
Then the function $S$, defines normal congruence of geodesics, and satisfies
the following non-linear partial differential equation
\begin{equation}
\frac{\partial S}{\partial \xi ^{1}}\frac{\partial S}{\partial \xi ^{2}}%
\frac{\partial S}{\partial \xi ^{3}}\frac{\partial S}{\partial \xi ^{4}}=%
\frac{\kappa ^{4}(\xi )}{4^{4}}\,.  \label{102}
\end{equation}
From the relation (\ref{100}) we obtain that
\begin{equation}
\left( V_{ind}\right) _{eu}=const\cdot \frac{1}{\kappa ^{4}}\,.  \label{103}
\end{equation}
Thus, the Lagrangian of the scalar field $S$ will have the form:
\begin{equation}
\mathfrak{L}=\frac{\partial S}{\partial \xi ^{1}}\frac{\partial S}{\partial
\xi ^{2}}\frac{\partial S}{\partial \xi ^{3}}\frac{\partial S}{\partial \xi
^{4}}\,.  \label{104}
\end{equation}
Correspondingly, the field equation will take the form  \small
\begin{equation}
\frac{\partial }{\partial \xi ^{1}}\left( \frac{\partial S}{\partial \xi ^{2}%
}\frac{\partial S}{\partial \xi ^{3}}\frac{\partial S}{\partial \xi ^{4}}%
\right) +\frac{\partial }{\partial \xi ^{2}}\left( \frac{\partial S}{%
\partial \xi ^{1}}\frac{\partial S}{\partial \xi ^{3}}\frac{\partial S}{%
\partial \xi ^{4}}\right) +\frac{\partial }{\partial \xi ^{3}}\left( \frac{%
\partial S}{\partial \xi ^{1}}\frac{\partial S}{\partial \xi ^{2}}\frac{%
\partial S}{\partial \xi ^{4}}\right) +\frac{\partial }{\partial \xi ^{4}}%
\left( \frac{\partial S}{\partial \xi ^{1}}\frac{\partial S}{\partial \xi
^{2}}\frac{\partial S}{\partial \xi ^{3}}\right) =0\,.  \label{105}
\end{equation}   \normalsize
Any function $S$, which depends on not all the coordinates $\xi ^{1},\xi ^{2},\xi
^{3},\xi ^{4}$ satisfies this equation.

Let the field $S$ depend on only one variable
\begin{equation}
s=\sqrt[4]{\xi ^{1}\xi ^{2}\xi ^{3}\xi ^{4}}\,,  \label{106}
\end{equation}
Substituting $S(s)$ into the field equation (\ref{105}) and using the
formula
\begin{equation}
\frac{\partial s}{\partial \xi ^{i}}=\frac{1}{4}\frac{s}{\xi ^{i}}\,,
\label{107}
\end{equation}
we obtain
\begin{equation}
\frac{d}{ds}\left( s\frac{dS}{ds}\right) =0\,.  \label{108}
\end{equation}
The same equation may be obtained easier, if the volume element
\begin{equation}
dV=\mathfrak{L}\,d\xi ^{1}d\xi ^{2}d\xi ^{3}d\xi ^{4}\,,  \label{109}
\end{equation}
is written, as a function of variable $s$ and three angular variables. After
integration of this element over the angles we obtain
\begin{equation}
dV_{s}=s^{3}\left( \frac{dS}{ds}\right) ^{4}\,ds\,.  \label{110}
\end{equation}
Via integration of the equation (\ref{108}), we get
\begin{equation}
S(s)=S_{0}\ln \frac{s}{s_{0}}\,,  \label{111}
\end{equation}
where $S_{0}$, $s_{0}$ are constants of integration, and also the relation
for the factor $\kappa $,
\begin{equation}
\kappa =\frac{|A|}{s}\,.  \label{112}
\end{equation}
This is quite interesting to compare the last two relations with the relations
(\ref{43}), (\ref{44}) and (\ref{60}), (\ref{61}).

Now we will find the trajectories of the motion of sample particles in the
four-dimensional Berwald-Moore space, if the function $S$, defining the congruence
of geodesics, has the form (\ref{111}), i.e. the factor satisfies the relation
(\ref{112}). The motion equations in this case will have the form
\begin{equation}
\dot{\xi}^{i}=\frac{\displaystyle\frac{\partial S}{\partial \xi ^{1}}\frac{%
\partial S}{\partial \xi ^{2}}\frac{\partial S}{\partial \xi ^{3}}\frac{%
\partial S}{\partial \xi ^{4}}}{\displaystyle\frac{\partial S}{\partial \xi
^{i}}}\lambda (\xi )\,,  \label{113}
\end{equation}
where $\lambda (\xi )\neq 0$ is a certain scalar function. Taking into
consideration the relation (\ref{107}) and via appropriate selection of $%
\lambda (\xi )$, motion equations may take a more simple form
\begin{equation}
\dot{\xi}^{i}=\xi ^{i}\,.  \label{114}
\end{equation}
Set the variable
\begin{equation}
x^{0}=\xi ^{1}+\xi ^{2}+\xi ^{3}+\xi ^{4}\,,  \label{115}
\end{equation}
which in the four-dimensional Berwald-Moore plays the same role as the
coordinate $x^{0}$ in the Minkowski space, then
\begin{equation}
\frac{d\xi ^{i}}{dx^{0}}=\frac{\xi ^{i}}{x^{0}}\qquad \Rightarrow \qquad \xi
^{i}=\xi _{0}^{i}\cdot x^{0}\,,  \label{116}
\end{equation}
where $\xi _{0}^{i}$ are constant. Thus, all the motion trajectories are
straight lines, passing through the origin, and the motion of sample bodies
will be uniform and rectilinear , with respect to the time variable $x^{0}$.

\section*{Conclusion}

The proposed new approach of the non-ambiguous construction of the field Lagrangians
basing on the metric function of the Finsler space requires that the fields which
are involved in the Lagrangian without their partial derivatives with respect to
coordinates,  {are expressed }via other fields so that these partial derivatives
over coordinates are involved in the Lagrangian, otherwise, this is not possible to
obtain the field equations as partial differential equations. Thus, the `art' of
Lagrangian construction is replaced with the `art' of representation of physical
fields using other fields.

For $n$-dimensional Riemannian or pseudo-Riemannian spaces with the metric tensor
$g_{ij}(x)$, the Lagrangian is given by
\[
\mathfrak{L}=\sqrt{|det(g_{ij}(x))|}\,.
\]
The metric tensor $g_{ij}(x)$ may be represented, for example, in the
following form:
\[
g_{ij}(x)=\sum\limits_{a=1}^{N}\varepsilon _{(a)}\frac{\partial f_{(a)}}{%
\partial x^{i}}\frac{\partial f_{(a)}}{\partial x^{j}}\,,
\]
here $\varepsilon _{(a)}=\pm 1$ are independent sign multiplicands, $%
f_{(a)}(x)$ are scalar functions, and $N\geq n$. If $N<n$, then $det\left(
g_{ij}(x)\right) =0$.

\normalsize

\end{document}